\def\be{\begin{equation}}
	\def\ee{\end{equation}}
\def\ba{\begin{array}}
	\def\ea{\end{array}}

\def\mathbi#1{\text{\em #1}}

\documentclass[prl,showpacs,twocolumn,amsmath]{revtex4}
\usepackage{amsmath}
\usepackage{amsfonts}
\usepackage{mathrsfs}
\usepackage{amssymb}
\usepackage{pifont}
\usepackage{epsfig,subfigure,dsfont,amsthm,amsbsy,mathrsfs,amscd}
\usepackage{epstopdf}
\usepackage{bbm}
\usepackage{color}
\def\qed{\leavevmode\unskip\penalty9999 \hbox{}\nobreak\hfill
	\quad\hbox{\leavevmode  \hbox to.77778em{%
			\hfil\vrule   \vbox to.675em%
			{\hrule width.6em\vfil\hrule}\vrule\hfil}}
	\par\vskip3pt}
\usepackage{leftidx}

\newtheorem{observation}{Observation}
\input amssym.def
\begin{document}
\title{\large\bf  Witness based nonlinear detection of quantum entanglement}

\author{Yiding Wang$^{1}$, Tinggui Zhang$^{1, \dag}$, Xiaofen Huang$^{1}$ and Shao-Ming Fei$^{2}$}
	\affiliation{ ${1}$ School of Mathematics and Statistics, Hainan Normal University, Haikou, 571158, China \\
		$2$ School of Mathematical Sciences, Capital Normal University, Beijing 100048, China \\
		$^{\dag}$ Correspondence to 050003@hainnu.edu.cn}
	
\bigskip
\bigskip
	
\begin{abstract}
We present a nonlinear entanglement detection strategy which detects entanglement that the linear detection strategy fails. We show that when the nonlinear entanglement detection strategy fails to detect the entanglement of an entangled state with two copies, it may succeed  with three or more copies. Based on our strategy, a witness combined with a suitable quanutm mechanical observable may detect the entanglement that can not be detected by the witness alone. Moreover, our strategy can also be applied to detect multipartite entanglement by using the witnesses for bipartite systems, as well as to entanglement concentrations.
\end{abstract}
	
 \maketitle
	
\section{I. Introduction}

One of the most distinctive features of quantum mechanics is the quantum correlations. The quantum entanglement \cite{rhph} is the most extensively studied quantum correlation, promising advantages over the classical correlations in many quantum information tasks such as quantum communications \cite{chb,cb,gr}, quantum computing \cite{ej,ad,mc}, quantum simulation \cite{sl} and quantum cryptography \cite{ake,grtz,lm}.

For the study of quantum entanglement, a basic issue is to determine whether a state is entangled or not. However, this problem is typically N-P hard \cite{lg}. Practical entanglement detection not only needs to ensure efficiency, that is to say use far fewer local measurements than those
needed for state tomography, but also has adequate detection power \cite{gtog,smmh}. The entanglement witnesses \cite{og} play important roles in experimental detection of quantum entanglement for unknown quantum states \cite{wszw,myht}. Entanglement witnesses are observables that are nonnegative for all separable states and negative for some entangled states, and have been widely studied in recent years \cite{jc,hq,fpvy,dmv,brlg,dcgs,mhr,arjs,ymzm,srlr,sxfl,atba,psw,ssm,zzsb,dhjs,vjfr,cgth,nyjj,msly,xsh,sdnb,dhjsk,nkhl,szt}. The authors in \cite{jc} show that each entanglement witness detecting the entanglement of a given bipartite state provides an estimation of the concurrence of this state. In \cite{hq} the authors provide two methods of constructing entanglement witnesses which detect entanglement that cannot be detected by the positive partial transpose (PPT) criterion and the realignment criterion. A series of new ultrafine entanglement witnesses have been proposed \cite{srlr,sxfl}, which can detect entangled states that the original entanglement witnesses cannot. The authors in \cite{ssm} obtain an upper bound on the entanglement witness function in the measurement-device-independent entanglement witness scenario. General and robust device-independent witnesses have been explored in \cite{zzsb}, which can be applied to identify entanglement in various arbitrary finite dimensional multipartite quantum systems, based merely on bipartite Bell inequalities. The author in \cite{xsh} presents a method to obtain the lower bounds of the concurrence, entanglement of formation and geometrical entanglement measure based on entanglement witnesses. In \cite{nkhl}, the authors extend the entanglement-certification toolbox from correlations in mutually unbiased bases to arbitrary bases, even without requiring aligned reference frames. Sun et al. \cite{szt} present an approach to estimate the operational distinguishability between an entangled state and any separable state directly from the measurement of an entanglement witness.

Nonlinear entanglement detection is currently the research focus \cite{vspw,llm,ymmb,tlhh,rh}, which can outperform linear counterparts \cite{ognl,mkmk,jmao,slly}. Liu et al. prove that if multi-copy joint measurements are allowed, the effectiveness of entanglement detection can be exponentially improved \cite{llm}. The authors in \cite{ymmb} characterize the genuine multipartite entanglement in the paradigm of multiple copies and conjecture a strict hierarchy of activatable states. Rico et al. introduce a systematic method for nonlinear entanglement detection based on trace polynomial inequalities \cite{rh}. In \cite{jmao}, the authors provide a method to construct nonlinear entanglement witnesses, which improve on linear entanglement witnesses in the sense that each non-linear entanglement witness detects more entangled states than its linear counterpart. Moreover, a nonlinear improvement of any entanglement witness for $2\times d$ quantum systems was introduced by \cite{slly}.

In fact, any entangled state can be detected by some proper entanglement witnesses, however, the construction of witnesses is not a straightforward work and relies on the specific structure of the state to be detected. Our motivation here is not to search for or construct new entangled witnesses, but to better utilize the existing entangled witnesses to improve their ability of entanglement detection. That is to say, given a set of entanglement witnesses, the question is if they can be still employed in a nonlinear strategy to detect entanglement in multiple copies of an unknown state, when the linear strategy fails. Our answer is positive. We prove that there are indeed witnesses for which the nonlinear strategy detects a different set of entangled states, while the linear counterparts fail to detect [Observation 1 and Observation 2]. Moreover, by combining witnesses with a positive semidefinite operator we show that there exist pairs of states and witnesses such that the linear detection fails, but the nonlinear detection succeeds. Furthermore, the entanglement detection ability can be improved by adjusting the positive semidefinite operators [Observation 3 and Example 4]. Our strategy may suggests a new perspective for detecting tripartite entangled states: using bipartite witnesses to measure two copies of the tripartite state [Observation 4 and Example 5]. The rest of this paper is organized as follows. In the second section, we provide our nonlinear detection strategy and the related applications. We summarize and discuss our conclusions in the last section.

\section{II. Witness based nonlinear detection of bipartite and multipartite entanglement}

An $n$-partite quantum state $\rho$ is said to be separable if it can be expressed as a convex combination of product states,
\begin{equation}\nonumber
\rho=\sum_{i}p_i\rho_i^1\otimes\rho_i^2\otimes...\otimes\rho_i^n,
\end{equation}
where $0\leq p_i\leq1$, $\sum_i p_i=1$, $\rho_i^k$ are the density matrices of the $k$th subsystem. Otherwise, $\rho$ is an entangled state. The entanglement witnesses provide powerful tools in detecting the entanglement of unknown quantum states. An entanglement witness is an observable $W$ such that $\text{Tr}(W\rho_{ent})<0$ for some entangled states $\rho_{ent}$ and $\text{Tr}(W\rho_{sep})\geq 0$ for all separable states $\rho_{sep}$. Therefore, $W$ gives rise to a hyperplane which separates the entangled states detected by $W$ from the convex set of separable states and the entangled states not detected by $W$. Such entanglement detection is called linear detection.
	
Nonlinear entanglement detection is given by witnesses based on nonlinear improvement of the linear ones \cite{ognl}, or by adopting multicopy scenarios \cite{ymmb,rh}. Recently in \cite{rh} the authors introduced the nonlinear entanglement witness by using tensor product of linear counterparts,
\begin{equation*}
\mathcal{W}=W_1\otimes W_2\otimes...\otimes W_n
\end{equation*}
such that $\text{Tr}(\mathcal{W}\rho_{ent}^{\otimes k})<0$ for some entangled states $\rho_{ent}$, while $\text{Tr}(\mathcal{W}\rho_{sep}^{\otimes k})\geq 0$ for all separable states, where $k$ depends on the local dimension and $n$. It is shown that there are two bipartite witnesses $W$ and $V$ and a bipartite entangled state $\rho_{AB}$ such that $\text{Tr}(W\rho_{AB})\geq0 $ and $\text{Tr}(V\rho_{AB})\geq 0$, but $\text{Tr}(W_{AA^{'}}\otimes V_{BB^{'}}\rho_{AB}^{\otimes 2})<0$. This approach elegantly improves the entanglement detection ability of the witnesses $W$ and $V$.

Different from the strategy given in \cite{rh}, we consider below another nonlinear entanglement detection strategy. Consider two copies of a bipartite state $\rho$, $\rho_{AB}$ and $\rho_{A^{'}B^{'}}$, and two bipartite witness $W$ and $V$. 
Bob measures the witness $V$ on systems $BA^{'}$, and Alice measures the witness $W$ on systems $AB^{'}$. The corresponding witness on the whole system is given by $\mathcal{W}=W_{AB^{'}}\otimes V_{BA^{'}}$. Then the expectation value of $\mathcal{W}$ is given by
$$
\langle\mathcal{W}\rangle_{\rho^{\otimes 2}}=\text{Tr}(\mathcal{W}(\rho_{AB}\otimes\rho_{A^{'}B^{'}})),
$$
which is nonnegative if $\rho_{AB}$ is separable. The strategy is illustrated in Fig. 1.
\begin{figure}[htbp]
\centering
\includegraphics[width=0.4\textwidth]{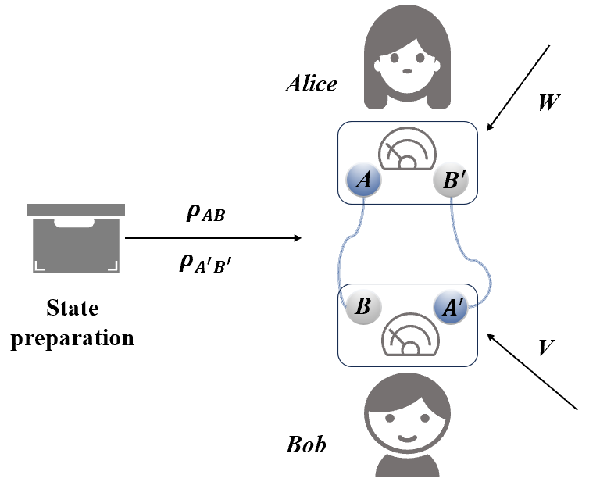}
\vspace{-1em} \caption{The expected value of $W_{AB^{'}}\otimes V_{BA^{'}}$ is obtained by measuring $\mathcal{W}$ on the subsystems $A$ and $B^{'}$, and $V$ on the subsystems $B$ and $A^{'}$.} \label{Fig.1}
\end{figure}

We have the following conclusion.

\begin{observation}
There exists an entangled state $\rho$ and witnesses $W$ and $V$ such that $Tr(W\rho)\geq 0$ and $Tr(V\rho)\geq 0$, but $Tr((W_{AB^{'}}\otimes V_{BA^{'}})\rho^{\otimes2})<0$.
\end{observation}

\mathbi{Example 1}. To illustrate the Observation 1 we consider the following 2-qubit witnesses based on the Bell state $|\phi^+\rangle=\frac{1}{\sqrt{2}}(|10\rangle+|01\rangle)$ and the standard Pauli operators $X$ and $Z$,
\begin{equation}\label{w1}
\begin{split}
W&=\mathbbm{1}-X\otimes X+Z\otimes Z,\\ V&=2|\phi^+\rangle\langle\phi^+|^{\tau_2},
\end{split}
\end{equation}
where $\tau_2$ stands for the partial transpose with respect to the second subsystem.

We first prove that $W$ and $V$ are qualified entanglement witnesses. 
$V$ is an entanglement witness since $|\phi^+\rangle$ is entangled \cite{phog}. 
Assume that there exists a separable state $\rho$ such that $\text{Tr}(W\rho)<0$. This means that $2(\rho_{11}+\rho_{44})<\rho_{14}+\rho_{14}^*+\rho_{23}+\rho_{23}^*=2(Re(\rho_{14})
+Re(\rho_{23}))$, namely, $|\rho_{14}|+|\rho_{23}|\geq Re(\rho_{14})+Re(\rho_{23})> \rho_{11}+\rho_{44}\geq2\sqrt{\rho_{11}\rho_{44}}$. Therefore, $|\rho_{23}|^2>\rho_{11}\rho_{44}$. According to positive partial transpose (PPT) criterion \cite{ap}, $\rho$ is entangled, which contradicts to the assumption. Hence, $\text{Tr}(W\rho)>0$ for all separable states $\rho$. Moreover, as $W$ has a negative eigenvalue of $-1$, $W$
is a well defined entanglement witness by definition \cite{bkml}.

To illustrate the Observation 1 we consider the Bell state $|\psi^+\rangle=\frac{1}{\sqrt{2}}(|00\rangle+|11\rangle)$ which gives important applications in such as quantum teleportation and quantum super dense coding \cite{mccyl,zjxjh,mhmp1,ygbhl}. By calculation, we have $\langle W\rangle_{\rho}=\langle V\rangle_{\rho}=1$ for the state $\rho=|\psi^+\rangle\langle\psi^+|$. That is to say, the entanglement of the state $\rho$ cannot be detected by either $W$ nor $V$. However, we have $\text{Tr}((W_{AB^{'}}\otimes V_{BA^{'}})\rho^{\otimes2})=-\frac{1}{2}$, see the detailed calculation in Appendix A. Thus the entanglement of $\rho$ is detected by $W$ and $V$ together.

Next we consider the detection of entanglement of states with non-vanishing imaginarity \cite{ahgg} by using real entanglement witnesses.

\mathbi{Example 2}. Let us consider the following quantum entangled state,
$$
\sigma=\frac{1}{2}\left(\begin{array}{cccc}
	0 & 0 & 0 & 0 \\
	0 & 1 & i & 0 \\
	0 & -i & 1 & 0 \\
	0 & 0 & 0 & 0 \\
\end{array}\right),
$$
where $i=\sqrt{-1}$. Here, due to hermitian properties, any real-valued linear witness cannot detect the entanglement of $\sigma$.  In fact, it can be verified that for any real valued entanglement witness $W_r$, $\text{Tr}(W_r\sigma)\geq 0$. Let us still use the real valued witnesses $W$ and $V$ given in Example 1. By calculation, we have $\text{Tr}(W\sigma)=0$ and $\text{Tr}(V\sigma)=1$. Using the strategy given in \cite{rh}, one obtains $\text{Tr}(W_{AA^{'}}\otimes V_{BB^{'}}\sigma^{\otimes2})=\frac{1}{2}$. However, our strategy shows that $\text{Tr}(W_{AB^{'}}\otimes V_{BA^{'}}\sigma^{\otimes2})=-\frac{1}{2}$, which detects the entanglement of $\sigma$, see the detailed calculations in Appendix B.

Example 2 shows that our strategy is complementary to the one given in \cite{rh}. Moreover,
when the entanglement can not be detected with two copies of a state, we may use multi-copies.

\begin{observation}
There exists an entangled state $\rho_{ent}$ and witnesses $W_i$, $i=1,2,3$, such that $Tr(W_i\rho_{ent})\geq0$ for $i=1,2,3$, and $Tr(W_{i,A_1B_2}\otimes W_{j,B_1A_2}\rho_{ent}^{\otimes2})\geq0$ for $i,j=1,2,3$, but
$$
Tr(W_{i,A_1B_2}\otimes W_{j,A_2B_3}\otimes W_{k,B_1A_3}\rho_{ent}^{\otimes3})<0
$$
for some $i,j,k=1,2,3$.
\end{observation}

\mathbi{Example 3}. The Werner state \cite{rfw},
$$
\rho_w=\frac{w\mathbbm{1}}{4}+(1-w)|\psi^+\rangle\langle\psi^+|,\,0\leq w\leq1,
$$
appears to be separable, entangled and nonlocal (it violates a Bell inequality)
at different parameter regions of $w$. Consider the following witnesses,
\begin{equation*}
\begin{split}
W_1&=\mathbbm{1}+X\otimes X-Y\otimes Y,\\
W_2&=2|\psi^-\rangle\langle\psi^-|^{\tau_2},\\
W_3&=2|\psi^+\rangle\langle\psi^+|^{\tau_2},
\end{split}
\end{equation*}
where $|\psi^-\rangle=\frac{1}{\sqrt{2}}(|00\rangle-|11\rangle)$.
We first show that $W_1$ is an entanglement witness. It can be verified that $W_1$ has a negative eigenvalue of $-1$. If $\text{Tr}(W_1\rho)<0$ for a separable state $\rho$, then $\rho_{11}+\rho_{22}+\rho_{33}+\rho_{44}+2(\rho_{14}+\rho_{14}^*)=1+4Re(\rho_{14})<0$. Since
	\begin{equation*}
		\frac{1}{4}(\rho_{11}+\rho_{44})^2\geq\rho_{11}\rho_{44}\geq|\rho_{14}|^2\geq Re(\rho_{14})^2>\frac{1}{16},
	\end{equation*}
we have $\rho_{11}+\rho_{44}>\frac{1}{2}$, and then
$2\sqrt{\rho_{22}\rho_{33}}\leq\rho_{22}+\rho_{33}<{1}/{2}$.
This implies that $\rho_{22}\rho_{33}<\frac{1}{16}<|\rho_{14}|^2$ and thus $\rho$ is entangled based on PPT criterion. Hence $\text{Tr}(W_1\rho)\geq 0$ for all separable states $\rho$.

According to PPT criterion, $\rho_w$ is entangled for $w<\frac{2}{3}$. The linear detection fails to detect the entanglement of $\rho_w$ as $\text{Tr}(W_i\rho_w)\geq0$, $i=1,2,3$. Moreover, the usual detection gives $\text{Tr}(W_{i,A_1A_2}\otimes W_{j,B_1B_2}\rho_w^{\otimes2})\geq0$ too. Our two copies detection does not work either, $\text{Tr}(W_{i,A_1B_2}\otimes W_{j,B_1A_2}\rho_w^{\otimes2})\geq0$ for $i,j=1,2,3$. Note that even with three copies, the usual ordering fails to detect the entanglement of $\rho_w$. Nevertheless, if we use three copies of $\rho_w^{\otimes3}$ based on our method, we obtain
$$
\text{Tr}(W_{1,A_1B_2}\otimes W_{2,A_2B_3}\otimes W_{3,B_1A_3}\rho_w^{\otimes3})<0
$$
for $0\leq w<0.206$. For instance, we have
$$
\text{Tr}(W_{1,A_1B_2}\otimes W_{2,A_2B_3}\otimes W_{3,B_1A_3}|\psi^+\rangle^{\otimes3})=-0.25<0
$$
for $w=0$, see Figure 2 and the details in Appendix C.
\begin{figure}[htbp]
\centering
\includegraphics[width=0.5\textwidth]{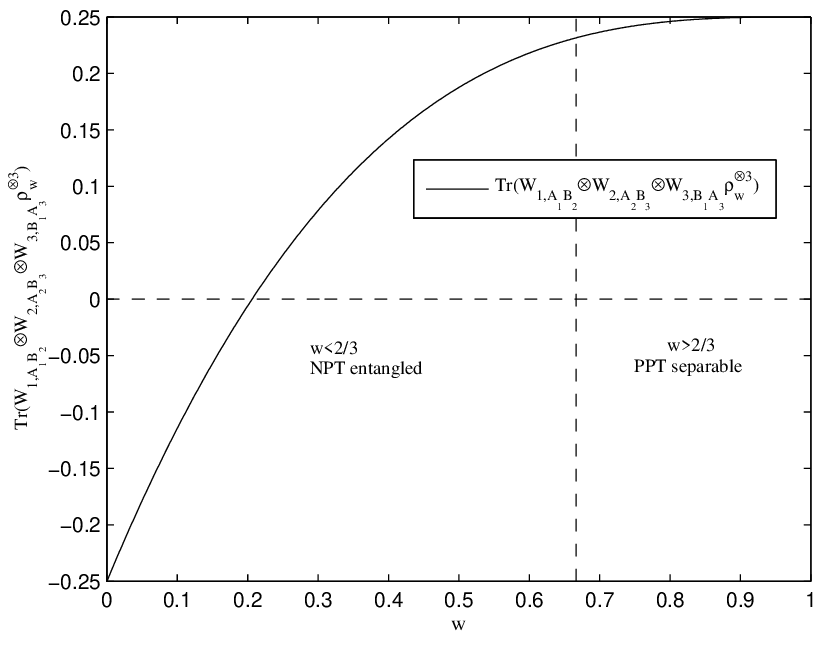}
\vspace{-1em} \caption{$\text{Tr}(W_{1,A_1B_2}\otimes W_{2,A_2B_3}\otimes W_{3,B_1A_3}\rho_w^{\otimes3})<0$ for $w<0.206$.} \label{Fig.2}
\end{figure}

Hence, our strategy applies to nonlinear detection of quantum entanglement involving general multi-copies of the quantum state $\rho$, see Figures 3 for the strategy with three and four copies by using the same or more new witnesses.
\begin{figure}[htbp]
	\centering
	\includegraphics[width=0.51\textwidth]{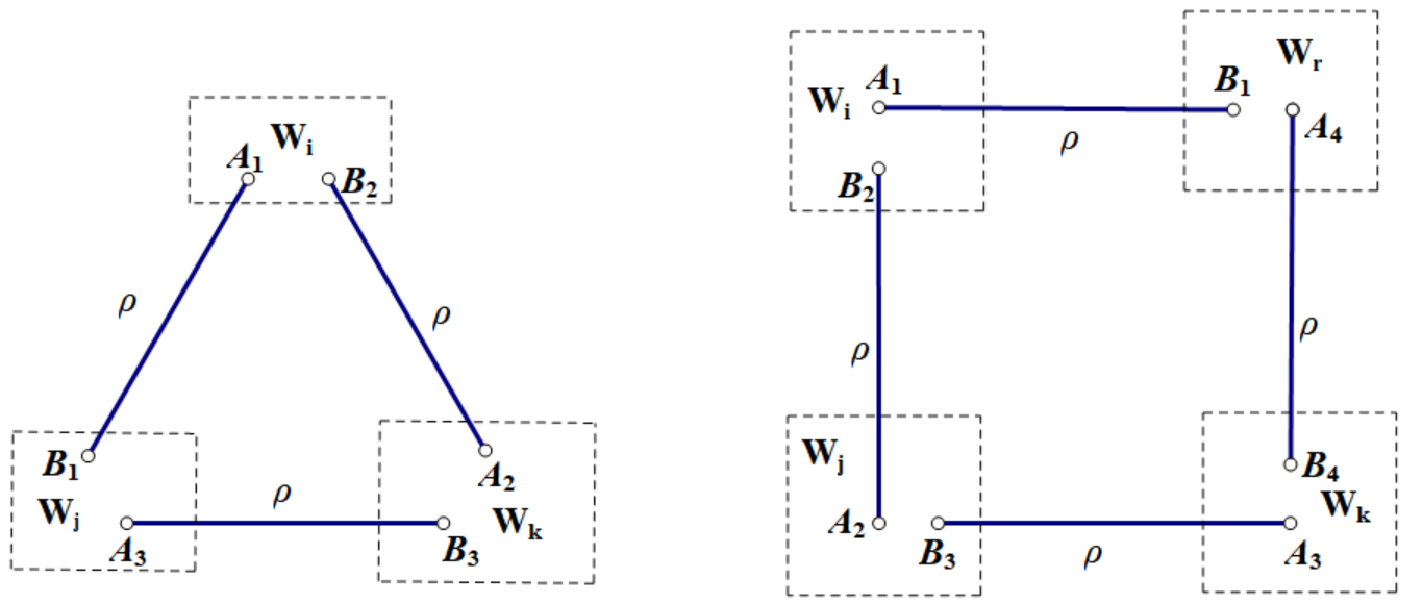}
	\vspace{-1em} \caption{The left (right) figure shows strategy of measuring witnesses on three (four) copies of state $\rho$.} \label{Fig.3}
\end{figure}

Moreover, in adopting more copies of a quantum state, although our strategy offers more new witnesses to be utilized, we may also use general observables instead of witnesses.Namely, one may replace some $W_i$ in $\mathcal{W}=W_1\otimes W_2\otimes...\otimes W_n$ with some positive semidefinite operators (not necessary entanglement witness). The expectation value will remain nonnegative for separable states. There exist entangled states $\rho_{ent}$ and witness $W$ such that the entanglement of $\rho_{ent}$ cannot be detected $W$, i.e., $\text{Tr}(W\rho_{ent})\geq0$. Rico et al. \cite{rh} showed that witnesses and positive semidefinite operator can detect entanglement in some states by using the usual ordering $A_1A_2|B_1B_2$. Based on the inspiration from Observation 1, a natural idea is the combination of witness and certain positive semidefinite operator $P$ to detect the entanglement by using our strategy $A_1B_2|B_1A_2$.

\begin{observation}
For some entangled states $\rho$ and witnesses $W$ such that $Tr(W\rho)\geq 0$, there exists positive semidefinite operators $P$ such that $Tr((P_{AB^{'}}\otimes W_{BA^{'}})\rho^{\otimes2})<0$.
\end{observation}

\mathbi{Example 4}. Let us consider the another Werner state,
$$
\rho_a=a|\psi^-\rangle\langle\psi^-|+\frac{1-a}{4}\mathbbm{1},
$$
where $0\leq a\leq1$. The state $\rho_a$ is entangled when $a>\frac{1}{3}$ according to the PPT criterion. We use the witness operator $W_3=2|\psi^+\rangle\langle\psi^+|^{\tau_2}$ and the positive semidefinite operator
$$
P=\left(\begin{array}{cccc}
	1 & 0 & 0 & -1 \\
	0 & 2 & -2 & 0 \\
	0 & -2 & 2 & 0 \\
	-1 & 0 & 0 & 1
\end{array}\right).
$$
Since $\text{Tr}(W_3\rho_a)=\frac{1+a}{2}\geq0$, $W_3$ fails to detect the entanglement of $\rho_a$. However, taking $\mathcal{W}=P\otimes W_3$ we obtain
$\text{Tr}(\mathcal{W}\rho_a^{\otimes2})<0$ for $a>\sqrt{\frac{3}{5}}$, see the detailed calculations in Appendix D. In fact, we can improve the result by replacing $P$ with
$$
P_b=\frac{1}{4b}\left(\begin{array}{cccc}
	1 & 0 & 0 & -1 \\
	0 & 2b & -2b & 0 \\
	0 & -2b & 2b & 0 \\
	-1 & 0 & 0 & 1
\end{array}\right).
$$
It is obvious that $P_b$ is a positive semidefinite operator for $b\geq1$. Let $\mathcal{W}_b=P_b\otimes W_3$. We have
$$
\text{Tr}(\mathcal{W}_b\rho_a^{\otimes2})=\frac{[(1-6b)a^2+2b+1]}{16b}.
$$
$\text{Tr}(\mathcal{W}_b\rho_a^{\otimes2})<0$ gives rise to $a>\sqrt{\frac{2b+1}{6b-1}}$ which approaches $\sqrt{\frac{1}{3}}$ when $b$ goes to $+\infty$. Namely, the entanglement of the state is detected for $a>\sqrt{\frac{1}{3}}$, see figures 4.
\begin{figure}[htbp]
	\centering
	\includegraphics[width=0.5\textwidth]{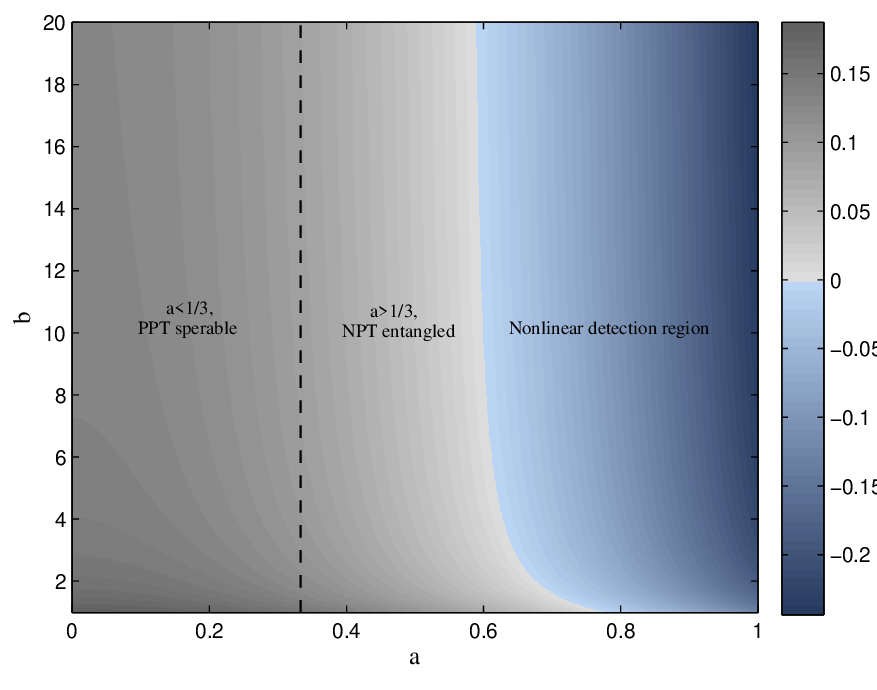}
	\vspace{-1em} \caption{$\text{Tr}(\mathcal{W}_b\rho_a^{\otimes2})$ versus parameters $a$ and $b$.} \label{Fig.4}
\end{figure}

Furthermore, besides bipartite states, our nonlinear entanglement detection strategy may be also applied to multi-copies of multipartite quantum states with a number of witness operators, as long as the dimension of the witness operators matches the dimension of the quantum states.

\begin{observation}
There exists some tripartite entangled state $\rho_{ent}$ and bipartite witnesses $W_i\,(i=1,2,3)$ such that $Tr(W_i\otimes W_j\otimes W_k\rho_{ent}^{\otimes2})\geq0$ $(i,j,k=1,2,3)$, but
\begin{equation*}
Tr(W_{i,AB^{'}}\otimes W_{j,BC^{'}}\otimes W_{k,CA^{'}}\rho_{ent}^{\otimes2})<0,
\end{equation*}
with some $i,j,k\in\{1,2,3\}.$
\end{observation}
Observation 4 may suggest a new perspective for detecting three-qubit entangled states: measure two-qubit witnesses on two copies of the three-qubit state, with dimensions matched each other. By using the witnesses $W_3=|\psi^+\rangle\langle\psi^+|^{\tau_2}$ and $W_4=\mathbbm{1}-Z\otimes Z-X\otimes X$ \cite{rh}, one can not successfully detect the entanglement of the Greenberger-Horne-Zeilinger (GHZ) state $|GHZ\rangle=1/\sqrt{2}(|000\rangle+|111\rangle)$ with two copies, under the measurement of
$W_{4,AA^{'}}\otimes W_{4,BB^{'}}\otimes W_{3,CC^{'}}$ \cite{rh}, because
$\text{Tr}(W_{4,AA^{'}}\otimes W_{4,BB^{'}}\otimes W_{3,CC^{'}}|GHZ\rangle^{\otimes2})\geq0$.
Nevertheless, we have
\begin{equation*}
\begin{split}
&\text{Tr}(W_{4,AA^{'}}\otimes W_{3,BB^{'}}\otimes W_{3,CC^{'}}|GHZ\rangle^{\otimes2})<0,\\
&\text{Tr}(W_{4,AB^{'}}\otimes W_{3,BC^{'}}\otimes W_{3,CA^{'}}|GHZ\rangle^{\otimes2})<0.
\end{split}
\end{equation*}
Similar consideration applies to the W states used in quantum computation and communication tasks \cite{xbzk,vngaa,tttw,xywyc}.

\mathbi{Example 5}. Consider the W state mixed with white noise,
	$$
	\rho_c=(1-c)|W\rangle\langle W|+\frac{c}{8}\mathbbm{1},
	$$
where $|W\rangle=1/\sqrt{3}(|001\rangle+|010\rangle+|100\rangle)$ and $c\in[0,1]$. We use witnesses $W_3=2|\psi^+\rangle\langle\psi^+|^{\tau_2}$ and $W_4=\mathbbm{1}-Z\otimes Z-X\otimes X$ above. It can be verified that
\begin{equation*}
\text{Tr}(W_i\otimes W_j\otimes W_k\rho_c^{\otimes2})\geq0,~\,i,j,k\in\{3,4\}.
\end{equation*}
However, by using our nonlinear detection, we have $\text{Tr}((W_{4,AB^{'}}\otimes W_{3,BC^{'}}\otimes W_{3,CA^{'}})\rho_c^{\otimes2})<0$ for $c<0.406$, namely, the entanglement is detected by $W_3$ and $W_4$, see the detailed calculations in Appendix E.

In \cite{mbmec} the authors proposed a witness for the W state,
\begin{equation*}
\mathcal{W}_W^{(1)}=\frac{2}{3}\mathbbm{1}-|W\rangle\langle W|.
\end{equation*}
It is shown that $\rho_c$ exhibits tripartite entanglement for $c<0.38$, see Figure 5. In other words, our nonlinear detection method can not only detect tripartite entanglement by using bipartite witnesses, but also improve the detection ability of tripartite witness. Note that Zhao et al. \cite{mjztg} introduced a tripartite entanglement criterion based on some complementary local observables, and proved that $\rho_c$ is entangled for $c<0.44$.
\begin{figure}[htbp]
	\centering
	\includegraphics[width=0.5\textwidth]{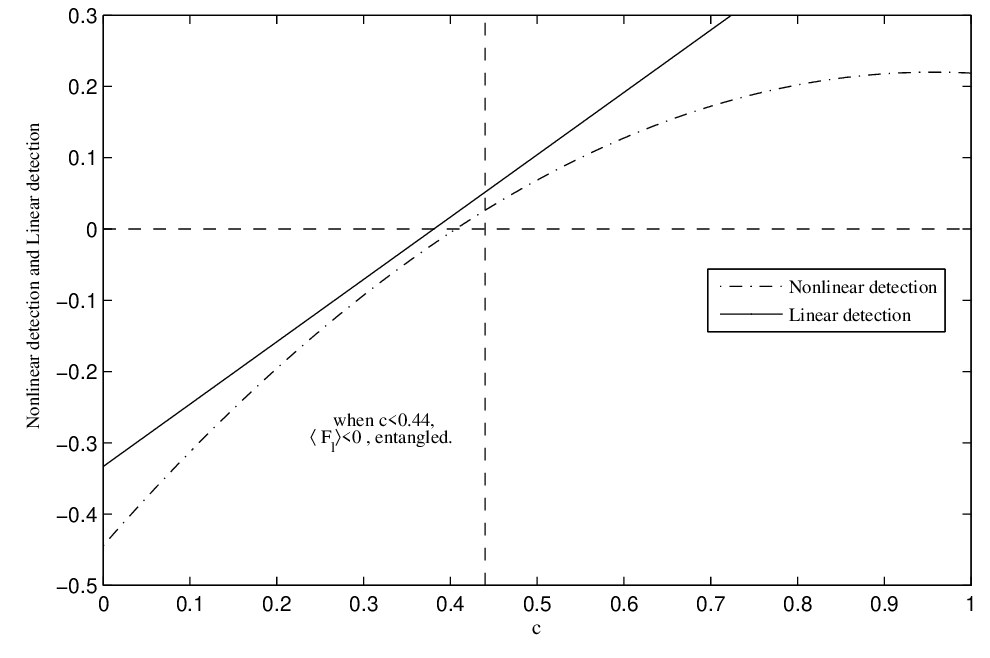}
	\vspace{-1em} \caption{The relations between nonlinear detection and linear detection for different values of $c$. It can be observed that $\text{Tr}((W_{4,AB^{'}}\otimes W_{3,BC^{'}}\otimes W_{3,CA^{'}})\rho_c^{\otimes2})<0$ for $c<0.406$ and $\text{Tr}(\mathcal{W}_W^{(1)}\rho_c)<0$ for $c<0.38$.} \label{Fig.5}
\end{figure}

Our nonlinear entanglement detection strategy can also be applied to entanglement concentration. This scenario can be applied to the following situation: the two copies of $|\varphi\rangle$ are shared by Alice and Bob in a way that Alice has $AB^{'}$ and Bob has $BA^{'}$. Bob's measurement concentrates the entanglement of the partition $AB$ and $A^{'}B^{'}$ into the Alice's subsystem $AB^{'}$. Furthermore, if appropriate measurements are taken, the final state on $AB^{'}$ can reach the maximally entangled state. Detailed scheme and steps are as follows. 

Consider two copies $|\varphi\rangle_{AB}$ and $|\varphi\rangle_{A^{'}B^{'}}$ of a bipartite pure state $|\varphi\rangle$ with full Schmidt rank. Generally $|\varphi\rangle$ can be expressed as
	$$|\varphi\rangle=\mathbbm{1}\otimes \Psi|\psi^+\rangle=\Psi^T\otimes\mathbbm{1}|\psi^+\rangle,$$
	where $\Psi$ is a $d\times d$ full ranked complex matrix such that $\text{Tr}(\Psi^{\dagger}\Psi)=1$ and $|\psi^+\rangle=\frac{1}{\sqrt{d}}\sum_{i=0}^{d-1}|ii\rangle$ is a Bell state. Performing a measurement $|m\rangle\langle m|$ on the subsystems $BA^{'}$ with $|m\rangle=\mathbbm{1}\otimes(\Psi^*)^{-1}|\psi^+\rangle$, i.e., a measurement $\mathbbm{1}_{AB^{'}}\otimes|m\rangle\langle m|_{BA^{'}}$ on $|\psi\rangle^{\otimes2}$, we obtain the final state $|\varphi\rangle$ of $AB^{'}$ with some probability,
\begin{equation*}
\frac{1}{p_m}\text{Tr}_{BA^{'}}(\mathbbm{1}_{AB^{'}}\otimes|m\rangle\langle m|_{BA^{'}}|\varphi\rangle\langle\varphi|^{\otimes2})\\
=\frac{1}{d^2p_m}|\varphi\rangle\langle\varphi|,
\end{equation*}
where $p_m=\text{Tr}(\mathbbm{1}_{AB^{'}}\otimes|m\rangle\langle m|_{BA^{'}}|\varphi\rangle\langle\varphi|^{\otimes2})$. Furthermore, if we replace the measurement acting on $BA^{'}$ with $|M\rangle\langle M|$ with
$|M\rangle=\mathbbm{1}\otimes(\Psi^*\Psi^*)^{-1}|\psi^+\rangle
=(\Psi^\dagger)^{-1}\otimes(\Psi^*)^{-1}|\psi^+\rangle$,
then we get the state
\begin{equation*}
\frac{1}{p_m^{'}}\text{Tr}_{BA^{'}}(\mathbbm{1}_{AB^{'}}\otimes|M\rangle\langle M|_{BA^{'}}|\varphi\rangle\langle\varphi|^{\otimes2})
=\frac{1}{d^2p_m^{'}}|\psi^+\rangle\langle\psi^+|,
\end{equation*}
where $p_m^{'}=\text{Tr}(\mathbbm{1}_{AB^{'}}\otimes|M\rangle\langle M|_{BA^{'}}|\varphi\rangle\langle\varphi|^{\otimes2})$, see the detailed derivation in Appendix F. Therefore, our strategy provides an effective approach of entanglement concentration.

\section{III. Conclusions and discussions}
In complementary to the one given in \cite{rh}, we have presented a different nonlinear entanglement detection strategy, which detects entanglement that the linear entanglement detection strategy fails. Meanwhile, we have generalized the strategy by adopting multi-copies and shown that when the detection for two copies fails, the detection for three or more copies can be successful, providing a new perspective for entanglement detection. Moreover, we have shown that there exist pairs of states and witnesses such that linear detection strategy fails, the nonlinear detection strategy given by combining the witness with a quantum mechanical observable succeeds, and the entanglement detection ability can be improved by adjusting the observables. Furthermore, our strategy can also be applied to entanglement concentrations.
Such nonlinear entanglement detection strategy is a new avenue. Our results may also highlight investigations on detections of other kind of quantum correlations.

\bigskip
{\bf Acknowledgments:} ~This work is supported by the National Natural Science Foundation of China (NSFC) under Grant Nos. 12204137, 12075159 and 12171044; the specific research fund of the Innovation Platform for Academicians of Hainan Province.

\appendix
\begin{widetext}
\section{Appendix A: Example 1}
Utilizing our strategy, we write down each term of $\rho_{AB}^{\otimes2}$ after measuring the operators $W_{AB^{'}}$ and $V_{BA^{'}}$. Without causing confusion, we abbreviate $|0\rangle_A|0\rangle_B|0\rangle_A^{'}|0\rangle_B^{'}$ as $|0000\rangle$, and similarly for other terms.
\begin{equation}\nonumber
\begin{split}
		 &|0000\rangle\langle0000|\xrightarrow[V_{BA^{'}}]{W_{AB^{'}}}2|0110\rangle\langle0000|-|1111\rangle\langle0000|,\hspace{1.15cm}|0000\rangle\langle0011|\xrightarrow[V_{BA^{'}}]{W_{AB^{'}}}2|0110\rangle\langle0011|-|1111\rangle\langle0011|,\\
		 &|0000\rangle\langle1100|\xrightarrow[V_{BA^{'}}]{W_{AB^{'}}}2|0110\rangle\langle1100|-|1111\rangle\langle1100|,\hspace{0.8cm}\bigstar|0000\rangle\langle1111|\xrightarrow[V_{BA^{'}}]{W_{AB^{'}}}2|0110\rangle\langle1111|-|1111\rangle\langle1111|,\\
		 &|0011\rangle\langle0000|\xrightarrow[V_{BA^{'}}]{W_{AB^{'}}}-|1010\rangle\langle0000|,\hspace{3.34cm}|0011\rangle\langle0011|\xrightarrow[V_{BA^{'}}]{W_{AB^{'}}}-|1010\rangle\langle0011|,\\
		 &|0011\rangle\langle1100|\xrightarrow[V_{BA^{'}}]{W_{AB^{'}}}-|1010\rangle\langle1100|,\hspace{3.34cm}|0011\rangle\langle1111|\xrightarrow[V_{BA^{'}}]{W_{AB^{'}}}-|1010\rangle\langle1111|,\\
		 &|1100\rangle\langle0000|\xrightarrow[V_{BA^{'}}]{W_{AB^{'}}}-|0101\rangle\langle0000|,\hspace{3.34cm}|1100\rangle\langle0011|\xrightarrow[V_{BA^{'}}]{W_{AB^{'}}}-|0101\rangle\langle0011|,\\
		 &|1100\rangle\langle1100|\xrightarrow[V_{BA^{'}}]{W_{AB^{'}}}-|0101\rangle\langle1100|,\hspace{3.34cm}|1100\rangle\langle1111|\xrightarrow[V_{BA^{'}}]{W_{AB^{'}}}-|0101\rangle\langle1111|,\\
		 \bigstar&|1111\rangle\langle0000|\xrightarrow[V_{BA^{'}}]{W_{AB^{'}}}2|1001\rangle\langle0000|-|0000\rangle\langle0000|,\hspace{1.15cm}|1111\rangle\langle0011|\xrightarrow[V_{BA^{'}}]{W_{AB^{'}}}2|1001\rangle\langle0011|-|0000\rangle\langle0011|,\\
		 &|1111\rangle\langle1100|\xrightarrow[V_{BA^{'}}]{W_{AB^{'}}}2|1001\rangle\langle1100|-|0000\rangle\langle1100|,\hspace{1.15cm}|1111\rangle\langle1111|\xrightarrow[V_{BA^{'}}]{W_{AB^{'}}}2|1001\rangle\langle1111|-|0000\rangle\langle1111|,\\
\end{split}
\end{equation}
There are only two elements appearing on the diagonal (marked with a pentagram). The corresponding coefficients are all $-\frac{1}{4}$. Thus we have $\text{Tr}((W_{AB^{'}}\otimes V_{BA^{'}})\rho_{AB}^{\otimes2})=-\frac{1}{4}+(-\frac{1}{4})=-\frac{1}{2}$.

\section{Appendix B: Example 2}
Using the strategy given in \cite{rh}, one obtains
\begin{equation}\nonumber
	\begin{split}
		 &|0101\rangle\langle0101|\xrightarrow[V_{BB^{'}}]{W_{AA^{'}}}2|0000\rangle\langle0101|-|1010\rangle\langle0101|,\hspace{0.8cm}|0110\rangle\langle0101|\xrightarrow[V_{BB^{'}}]{W_{AA^{'}}}-|1100\rangle\langle0101|,\\
		 &|0101\rangle\langle0110|\xrightarrow[V_{BB^{'}}]{W_{AA^{'}}}2|0000\rangle\langle0110|-|1010\rangle\langle0110|,\hspace{0.8cm}|0110\rangle\langle0110|\xrightarrow[V_{BB^{'}}]{W_{AA^{'}}}-|1100\rangle\langle0110|,\\
		 &|0101\rangle\langle1001|\xrightarrow[V_{BB^{'}}]{W_{AA^{'}}}2|0000\rangle\langle1001|-|1010\rangle\langle1001|,\hspace{0.8cm}|0110\rangle\langle1001|\xrightarrow[V_{BB^{'}}]{W_{AA^{'}}}-|1100\rangle\langle1001|,\\
		 \bigstar&|0101\rangle\langle1010|\xrightarrow[V_{BB^{'}}]{W_{AA^{'}}}2|0000\rangle\langle1010|-|1010\rangle\langle1010|,\hspace{0.8cm}|0110\rangle\langle1010|\xrightarrow[V_{BB^{'}}]{W_{AA^{'}}}-|1100\rangle\langle1010|,\\
		 \bigstar&|1010\rangle\langle0101|\xrightarrow[V_{BB^{'}}]{W_{AA^{'}}}2|1111\rangle\langle0101|-|0101\rangle\langle0101|,\hspace{0.8cm}|1001\rangle\langle0101|\xrightarrow[V_{BB^{'}}]{W_{AA^{'}}}-|0011\rangle\langle0101|,\\
		 &|1010\rangle\langle0110|\xrightarrow[V_{BB^{'}}]{W_{AA^{'}}}2|1111\rangle\langle0110|-|0101\rangle\langle0110|,\hspace{0.8cm}|1001\rangle\langle0110|\xrightarrow[V_{BB^{'}}]{W_{AA^{'}}}-|0011\rangle\langle0110|,\\
		 &|1010\rangle\langle1001|\xrightarrow[V_{BB^{'}}]{W_{AA^{'}}}2|1111\rangle\langle1001|-|0101\rangle\langle1001|,\hspace{0.8cm}|1001\rangle\langle1001|\xrightarrow[V_{BB^{'}}]{W_{AA^{'}}}-|0011\rangle\langle1001|,\\
		 &|1010\rangle\langle1010|\xrightarrow[V_{BB^{'}}]{W_{AA^{'}}}2|1111\rangle\langle1010|-|0101\rangle\langle1010|,\hspace{0.8cm}|1001\rangle\langle1010|\xrightarrow[V_{BB^{'}}]{W_{AA^{'}}}-|0011\rangle\langle1010|.
\end{split}
\end{equation}
So we have $$\text{Tr}(W_{AA^{'}}\otimes V_{BB^{'}}\sigma^{\otimes2})=-(\frac{i}{2})^2-(-\frac{i}{2})^2=\frac{1}{4}+\frac{1}{4}=\frac{1}{2}.$$
However, our nonlinear strategy gives
\begin{equation}\nonumber
	\begin{split}
		 &|0110\rangle\langle0101|\xrightarrow[V_{BA^{'}}]{W_{AB^{'}}}2|0000\rangle\langle0101|-|1001\rangle\langle0101|,\hspace{0.8cm}|0101\rangle\langle0101|\xrightarrow[V_{BA^{'}}]{W_{AB^{'}}}-|1100\rangle\langle0101|,\\
		 \bigstar&|0110\rangle\langle1001|\xrightarrow[V_{BA^{'}}]{W_{AB^{'}}}2|0000\rangle\langle1001|-|1001\rangle\langle1001|,\hspace{0.8cm}|0101\rangle\langle0110|\xrightarrow[V_{BA^{'}}]{W_{AB^{'}}}-|1100\rangle\langle0110|,\\
		 &|0110\rangle\langle0110|\xrightarrow[V_{BA^{'}}]{W_{AB^{'}}}2|0000\rangle\langle0110|-|1001\rangle\langle0110|,\hspace{0.8cm}|0101\rangle\langle1001|\xrightarrow[V_{BA^{'}}]{W_{AB^{'}}}-|1100\rangle\langle1001|,\\
		 &|0110\rangle\langle1010|\xrightarrow[V_{BA^{'}}]{W_{AB^{'}}}2|0000\rangle\langle1010|-|1001\rangle\langle1010|,\hspace{0.8cm}|0101\rangle\langle1010|\xrightarrow[V_{BA^{'}}]{W_{AB^{'}}}-|1100\rangle\langle1010|,\\
		 &|1001\rangle\langle0101|\xrightarrow[V_{BA^{'}}]{W_{AB^{'}}}2|1111\rangle\langle0101|-|0110\rangle\langle0101|,\hspace{0.8cm}|1010\rangle\langle0101|\xrightarrow[V_{BA^{'}}]{W_{AB^{'}}}-|0011\rangle\langle0101|,\\
		 \bigstar&|1001\rangle\langle0110|\xrightarrow[V_{BA^{'}}]{W_{AB^{'}}}2|1111\rangle\langle0110|-|0110\rangle\langle0110|,\hspace{0.8cm}|1010\rangle\langle0110|\xrightarrow[V_{BA^{'}}]{W_{AB^{'}}}-|0011\rangle\langle0110|,\\
		 &|1001\rangle\langle1001|\xrightarrow[V_{BA^{'}}]{W_{AB^{'}}}2|1111\rangle\langle1001|-|0110\rangle\langle1001|,\hspace{0.8cm}|1010\rangle\langle1001|\xrightarrow[V_{BA^{'}}]{W_{AB^{'}}}-|0011\rangle\langle1001|,\\
		 &|1001\rangle\langle1010|\xrightarrow[V_{BA^{'}}]{W_{AB^{'}}}2|1111\rangle\langle1010|-|0110\rangle\langle1010|,\hspace{0.8cm}|1010\rangle\langle1010|\xrightarrow[V_{BA^{'}}]{W_{AB^{'}}}-|0011\rangle\langle1010|.\\
	\end{split}
\end{equation}
Therefore, $$\text{Tr}(W_{AB^{'}}\otimes V_{BA^{'}}\sigma^{\otimes2})=-(\frac{i}{2}.\frac{-i}{2})-(\frac{-i}{2}.\frac{i}{2})=-\frac{1}{4}-\frac{1}{4}=-\frac{1}{2}$$.

\section{Appendix C: Example 3}
We calculate the value of $\text{Tr}(W_{1,A_1B_2}\otimes W_{2,A_2B_3}\otimes W_{2,B_1A_3}\rho_w^{\otimes3})$. For simplicity, we only list the terms that contribute to the trace,
\begin{equation*}
	\begin{split}
		 &|000000\rangle\langle000000|\xrightarrow[W_{2,A_2B_3},W_{3,B_1A_3}]{W_{1,A_1B_2}}|000000\rangle\langle000000|+2|100100\rangle\langle000000|,\\
		 &|000011\rangle\langle111100|\xrightarrow[W_{2,A_2B_3},W_{3,B_1A_3}]{W_{1,A_1B_2}}-|011000\rangle\langle111100|-2|111100\rangle\langle111100|,\\
		 &|111100\rangle\langle000011|\xrightarrow[W_{2,A_2B_3},W_{3,B_1A_3}]{W_{1,A_1B_2}}-|100111\rangle\langle000011|-2|000011\rangle\langle000011|,\\
		 &|111111\rangle\langle111111|\xrightarrow[W_{2,A_2B_3},W_{3,B_1A_3}]{W_{1,A_1B_2}}|111111\rangle\langle111111|+2|011011\rangle\langle111111|.\\
		 &|001001\rangle\langle001001|\xrightarrow[W_{2,A_2B_3},W_{3,B_1A_3}]{W_{1,A_1B_2}}|001001\rangle\langle001001|+2|101101\rangle\langle001001|,\\
		 &|010010\rangle\langle010010|\xrightarrow[W_{2,A_2B_3},W_{3,B_1A_3}]{W_{1,A_1B_2}}|010010\rangle\langle010010|+2|110110\rangle\langle010010|,\\
		 &|011011\rangle\langle011011|\xrightarrow[W_{2,A_2B_3},W_{3,B_1A_3}]{W_{1,A_1B_2}}|011011\rangle\langle011011|+2|111111\rangle\langle011011|,\\
		 &|100100\rangle\langle100100|\xrightarrow[W_{2,A_2B_3},W_{3,B_1A_3}]{W_{1,A_1B_2}}|100100\rangle\langle100100|+2|000000\rangle\langle100100|,\\
		 &|101101\rangle\langle101101|\xrightarrow[W_{2,A_2B_3},W_{3,B_1A_3}]{W_{1,A_1B_2}}|101101\rangle\langle101101|+2|001001\rangle\langle101101|,\\
		 &|110110\rangle\langle110110|\xrightarrow[W_{2,A_2B_3},W_{3,B_1A_3}]{W_{1,A_1B_2}}|110110\rangle\langle110110|+2|010010\rangle\langle110110|,\\
		 &|000100\rangle\langle000100|\xrightarrow[W_{2,A_2B_3},W_{3,B_1A_3}]{W_{1,A_1B_2}}|000100\rangle\langle000100|,\hspace{0.5cm}|001101\rangle\langle001101|\xrightarrow[W_{2,A_2B_3},W_{3,B_1A_3}]{W_{1,A_1B_2}}|001101\rangle\langle001101|,\\
	\end{split}
\end{equation*}

\begin{equation}
\begin{split}
&|010110\rangle\langle010110|\xrightarrow[W_{2,A_2B_3},W_{3,B_1A_3}]{W_{1,A_1B_2}}|010110\rangle\langle010110|,\hspace{0.5cm}|011111\rangle\langle011111|\xrightarrow[W_{2,A_2B_3},W_{3,B_1A_3}]{W_{1,A_1B_2}}|011111\rangle\langle011111|,\\
&|111011\rangle\langle111011|\xrightarrow[W_{2,A_2B_3},W_{3,B_1A_3}]{W_{1,A_1B_2}}|111011\rangle\langle111011|,\hspace{0.5cm}|110010\rangle\langle110010|\xrightarrow[W_{2,A_2B_3},W_{3,B_1A_3}]{W_{1,A_1B_2}}|110010\rangle\langle110010|,\\
&|101001\rangle\langle101001|\xrightarrow[W_{2,A_2B_3},W_{3,B_1A_3}]{W_{1,A_1B_2}}|101001\rangle\langle101001|,\hspace{0.5cm}|100000\rangle\langle100000|\xrightarrow[W_{2,A_2B_3},W_{3,B_1A_3}]{W_{1,A_1B_2}}|100000\rangle\langle100000|,\\
\end{split}
\end{equation}
Therefore,
$$
\text{Tr}(W_{1,A_1B_2}\otimes W_{2,A_2B_3}\otimes W_{3,B_1A_3}\rho_w^{\otimes3})=2(\frac{2-w}{4})^3-4(\frac{1-w}{2})^3
+6\frac{w(2-w)^2}{64}+6\frac{w^2(2-w)}{64}+2\frac{w^3}{64},
$$
from which it is derived that $\text{Tr}(W_{1,A_1B_2}\otimes W_{2,A_2B_3}\otimes W_{3,B_1A_3}\rho_w^{\otimes3})<0$ when $w<0.206$. In particular, when $w=0$ we have
$$
\text{Tr}(W_{1,A_1B_2}\otimes W_{2,A_2B_3}\otimes W_{3,B_1A_3}|\psi^+\rangle^{\otimes3})=-0.25<0.
$$

\section{Appendix D: Example 4}

Concerning the calculation of $\text{Tr}(\mathcal{W}\rho_a^{\otimes2})$, we have
\begin{equation*}
\begin{split}
\bigstar&|0000\rangle\langle0000|\xrightarrow[W_{BA^{'}}]{P_{AB^{'}}}|0000\rangle\langle0000|-|1001\rangle\langle0000|,\hspace{1.5cm}|0100\rangle\langle0100|\xrightarrow[W_{BA^{'}}]{P_{AB^{'}}}|0010\rangle\langle0100|-|1011\rangle\langle0100|,\\
&|0000\rangle\langle0011|\xrightarrow[W_{BA^{'}}]{P_{AB^{'}}}|0000\rangle\langle0011|-|1001\rangle\langle0011|,\hspace{1.5cm}|0100\rangle\langle0111|\xrightarrow[W_{BA^{'}}]{P_{AB^{'}}}|0010\rangle\langle0111|-|1011\rangle\langle0111|,\\
&|0000\rangle\langle1100|\xrightarrow[W_{BA^{'}}]{P_{AB^{'}}}|0000\rangle\langle1100|-|1001\rangle\langle1100|,\hspace{1.5cm}|0101\rangle\langle0101|\xrightarrow[W_{BA^{'}}]{P_{AB^{'}}}2|0011\rangle\langle0101|-2|1010\rangle\langle0101|,\\
&|0000\rangle\langle1111|\xrightarrow[W_{BA^{'}}]{P_{AB^{'}}}|0000\rangle\langle1111|-|1001\rangle\langle1111|,\hspace{1.15cm}\bigstar|0110\rangle\langle0110|\xrightarrow[W_{BA^{'}}]{P_{AB^{'}}}|0110\rangle\langle0110|-|1111\rangle\langle0110|,\\
&|0011\rangle\langle0000|\xrightarrow[W_{BA^{'}}]{P_{AB^{'}}}2|0101\rangle\langle0000|-2|1100\rangle\langle0000|,\hspace{1.12cm}|0111\rangle\langle0100|\xrightarrow[W_{BA^{'}}]{P_{AB^{'}}}2|0111\rangle\langle0100|-2|1110\rangle\langle0100|,\\
&|0011\rangle\langle0011|\xrightarrow[W_{BA^{'}}]{P_{AB^{'}}}2|0101\rangle\langle0011|-2|1100\rangle\langle0011|,\hspace{0.77cm}\bigstar|0111\rangle\langle0111|\xrightarrow[W_{BA^{'}}]{P_{AB^{'}}}2|0111\rangle\langle0111|-2|1110\rangle\langle0111|,\\
\bigstar&|0011\rangle\langle1100|\xrightarrow[W_{BA^{'}}]{P_{AB^{'}}}2|0101\rangle\langle1100|-2|1100\rangle\langle1100|,\hspace{0.77cm}\bigstar|1000\rangle\langle1000|\xrightarrow[W_{BA^{'}}]{P_{AB^{'}}}2|1000\rangle\langle1000|-2|0001\rangle\langle1000|,\\
&|0011\rangle\langle1111|\xrightarrow[W_{BA^{'}}]{P_{AB^{'}}}2|0101\rangle\langle1111|-2|1100\rangle\langle1111|,\hspace{1.12cm}|1000\rangle\langle1011|\xrightarrow[W_{BA^{'}}]{P_{AB^{'}}}2|1000\rangle\langle1011|-2|0001\rangle\langle1011|,\\
&|1100\rangle\langle0000|\xrightarrow[W_{BA^{'}}]{P_{AB^{'}}}2|1010\rangle\langle0000|-2|0011\rangle\langle0000|,\hspace{0.77cm}\bigstar|1001\rangle\langle1001|\xrightarrow[W_{BA^{'}}]{P_{AB^{'}}}|1001\rangle\langle1001|-|0000\rangle\langle1001|,\\
\bigstar&|1100\rangle\langle0011|\xrightarrow[W_{BA^{'}}]{P_{AB^{'}}}2|1010\rangle\langle0011|-2|0011\rangle\langle0011|,\hspace{1.12cm}|1010\rangle\langle1010|\xrightarrow[W_{BA^{'}}]{P_{AB^{'}}}2|1100\rangle\langle1010|-2|0101\rangle\langle1010|,\\
\end{split}
\end{equation*}

\begin{equation}
\begin{split}
&|1100\rangle\langle1100|\xrightarrow[W_{BA^{'}}]{P_{AB^{'}}}2|1010\rangle\langle1100|-2|0011\rangle\langle1100|,\hspace{1.12cm}|1011\rangle\langle1000|\xrightarrow[W_{BA^{'}}]{P_{AB^{'}}}|1101\rangle\langle1000|-|0100\rangle\langle1000|,\\
&|1100\rangle\langle1111|\xrightarrow[W_{BA^{'}}]{P_{AB^{'}}}2|1010\rangle\langle1111|-2|0011\rangle\langle1111|,\hspace{1.12cm}|1011\rangle\langle1011|\xrightarrow[W_{BA^{'}}]{P_{AB^{'}}}|1101\rangle\langle1011|-|0100\rangle\langle1011|,\\
&|1111\rangle\langle0000|\xrightarrow[W_{BA^{'}}]{P_{AB^{'}}}|1111\rangle\langle0000|-|0110\rangle\langle0000|,\hspace{1.5cm}|1101\rangle\langle0001|\xrightarrow[W_{BA^{'}}]{P_{AB^{'}}}|1011\rangle\langle0001|-|0010\rangle\langle0001|,\\
&|1111\rangle\langle0011|\xrightarrow[W_{BA^{'}}]{P_{AB^{'}}}|1111\rangle\langle0011|-|0110\rangle\langle0011|,\hspace{1.5cm}|1110\rangle\langle0010|\xrightarrow[W_{BA^{'}}]{P_{AB^{'}}}2|1110\rangle\langle0010|-2|0111\rangle\langle0010|,\\
&|1111\rangle\langle1100|\xrightarrow[W_{BA^{'}}]{P_{AB^{'}}}|1111\rangle\langle1100|-|0110\rangle\langle1100|,\hspace{1.15cm}\bigstar|1110\rangle\langle1110|\xrightarrow[W_{BA^{'}}]{P_{AB^{'}}}2|1110\rangle\langle1110|-2|0111\rangle\langle1110|,\\
\bigstar&|1111\rangle\langle1111|\xrightarrow[W_{BA^{'}}]{P_{AB^{'}}}|1111\rangle\langle1111|-|0110\rangle\langle1111|,\hspace{1.5cm}|1101\rangle\langle1101|\xrightarrow[W_{BA^{'}}]{P_{AB^{'}}}|1011\rangle\langle1101|-|0010\rangle\langle1101|,\\
\bigstar&|0001\rangle\langle0001|\xrightarrow[W_{BA^{'}}]{P_{AB^{'}}}2|0001\rangle\langle0001|-2|1000\rangle\langle0001|,\hspace{1.12cm}|0010\rangle\langle0010|\xrightarrow[W_{BA^{'}}]{P_{AB^{'}}}|0100\rangle\langle0010|-|1101\rangle\langle0010|,\\
&|0001\rangle\langle1101|\xrightarrow[W_{BA^{'}}]
{P_{AB^{'}}}2|0001\rangle\langle1101|-2|1000\rangle\langle1101|,
\hspace{1.12cm}|0010\rangle\langle1110|
\xrightarrow[W_{BA^{'}}]{P_{AB^{'}}}|0100\rangle\langle1110|-|1101\rangle\langle1110|,
\end{split}
\end{equation}
where the elements that contribute to the trace have been marked by a pentagram.
Then we have
$$
\text{Tr}(\mathcal{W}\rho_a^{\otimes2})=\frac{(1+a)^2}{8}-a^2+\frac{1-a^2}{2}+\frac{(1-a)^2}{8}=\frac{3-5a^2}{4}.
$$

\section{Appendix E: Example 5}
We first show that
	\begin{equation*}
	\text{Tr}(W_i\otimes W_j\otimes W_k\rho_c^{\otimes2})\geq0,\,i,j,k\in\{3,4\}.
\end{equation*}

\begin{equation*}
	\begin{split}
		&\text{Tr}(W_3\otimes W_3\otimes W_3\rho_c^{\otimes2})=(\frac{1}{3}-\frac{c}{3})(\frac{4}{3}-\frac{c}{3})+(\frac{1}{3}-\frac{5c}{24})(\frac{1}{3}+\frac{c}{24})+\frac{5c^2}{64}\geq0,\\
		&\text{Tr}(W_3\otimes W_3\otimes W_4\rho_c^{\otimes2})=(\frac{1}{3}-\frac{c}{3})(\frac{4}{3}-\frac{c}{3})+c(\frac{1}{3}-\frac{7c}{48})+\frac{c^2}{16}\geq0,\\
		&\text{Tr}(W_3\otimes W_4\otimes W_3\rho_c^{\otimes2})=2(\frac{1}{3}-\frac{c}{3})(\frac{4}{3}-\frac{c}{3})+c(\frac{1}{3}-\frac{5c}{24})+\frac{c^2}{8}\geq0,\\
		&\text{Tr}(W_3\otimes W_4\otimes W_4\rho_c^{\otimes2})=(\frac{1}{3}-\frac{5c}{24})(\frac{4}{3}-\frac{c}{3})+\frac{c(1-c)}{2}+c(\frac{1}{3}-\frac{5c}{24})+\frac{c^2}{4}\geq0,\\
		&\text{Tr}(W_4\otimes W_3\otimes W_3\rho_c^{\otimes2})=(\frac{1}{3}-\frac{c}{3})(\frac{4}{3}-\frac{c}{3})+\frac{c}{2}(\frac{2}{3}-\frac{7c}{24})+\frac{c^2}{16}\geq0,\\
		&\text{Tr}(W_4\otimes W_3\otimes W_4\rho_c^{\otimes2})=4[(\frac{1}{3}-\frac{c}{3})+2(\frac{1}{3}-\frac{5c}{24})]^2+\frac{c^2}{4}\geq0,\\
		&\text{Tr}(W_4\otimes W_4\otimes W_3\rho_c^{\otimes2})=(\frac{1}{3}-\frac{5c}{24})(\frac{4}{3}-\frac{c}{3})+c(\frac{1}{3}-\frac{5c}{24})+\frac{c(1-c)}{2}+\frac{c^2}{4}\geq0,\\
		&\text{Tr}(W_4\otimes W_4\otimes W_4\rho_c^{\otimes2})=c(\frac{4}{3}-\frac{c}{3})\geq0,\\
	\end{split}
\end{equation*}
With respect to the calculation of $\text{Tr}((W_{4,AB^{'}}\otimes W_{3,BC^{'}}\otimes W_{3,CA^{'}})\rho_c^{\otimes2})$, we only list the terms that contribute to the trace for simplicity,

	\begin{equation}\nonumber
		\begin{split}
			 &|000010\rangle\langle000010|\xrightarrow[W_{3,BC^{'}},W_{3,CA^{'}}]{W_{4,AB^{'}}}2|000010\rangle\langle000010|-|100000\rangle\langle000010|,\\
			 &|001010\rangle\langle100100|\xrightarrow[W_{3,BC^{'}},W_{3,CA^{'}}]{W_{4,AB^{'}}}2|000110\rangle\langle100100|-|100100\rangle\langle100100|,\\
			 &|001110\rangle\langle001110|\xrightarrow[W_{3,BC^{'}},W_{3,CA^{'}}]{W_{4,AB^{'}}}2|001110\rangle\langle001110|-|101100\rangle\langle001110|,\\
			 &|010010\rangle\langle100001|\xrightarrow[W_{3,BC^{'}},W_{3,CA^{'}}]{W_{4,AB^{'}}}2|000011\rangle\langle100001|-|100001\rangle\langle100001|,\\
			 &|010011\rangle\langle010011|\xrightarrow[W_{3,BC^{'}},W_{3,CA^{'}}]{W_{4,AB^{'}}}2|010011\rangle\langle010011|-|110001\rangle\langle010011|,\\
			 &|011111\rangle\langle011111|\xrightarrow[W_{3,BC^{'}},W_{3,CA^{'}}]{W_{4,AB^{'}}}2|011111\rangle\langle011111|-|111101\rangle\langle011111|,\\
			 &|100000\rangle\langle100000|\xrightarrow[W_{3,BC^{'}},W_{3,CA^{'}}]{W_{4,AB^{'}}}2|100000\rangle\langle100000|-|000010\rangle\langle100000|,\\
			 &|100001\rangle\langle010010|\xrightarrow[W_{3,BC^{'}},W_{3,CA^{'}}]{W_{4,AB^{'}}}2|110000\rangle\langle010010|-|010010\rangle\langle010010|,\\
			 &|100100\rangle\langle001010|\xrightarrow[W_{3,BC^{'}},W_{3,CA^{'}}]{W_{4,AB^{'}}}2|101000\rangle\langle001010|-|001010\rangle\langle001010|,\\
			 &|101100\rangle\langle101100|\xrightarrow[W_{3,BC^{'}},W_{3,CA^{'}}]{W_{4,AB^{'}}}2|101100\rangle\langle101100|-|001110\rangle\langle101100|,\\
			 &|110001\rangle\langle110001|\xrightarrow[W_{3,BC^{'}},W_{3,CA^{'}}]{W_{4,AB^{'}}}2|110001\rangle\langle110001|-|010011\rangle\langle110001|,\\
			 &|111101\rangle\langle111101|\xrightarrow[W_{3,BC^{'}},W_{3,CA^{'}}]{W_{4,AB^{'}}}2|111101\rangle\langle111101|-|011111\rangle\langle111101|.
		\end{split}
	\end{equation}
 Hence, we have $$\text{Tr}((W_{4,AB^{'}}\otimes W_{3,BC^{'}}\otimes W_{3,CA^{'}})\rho_c^{\otimes2})=2.(\frac{c}{8})^2+12.\frac{c}{8}.\frac{8-5c}{24}+4.(\frac{1-c}{3})^2,$$
 where $\text{Tr}((W_{4,AB^{'}}\otimes W_{3,BC^{'}}\otimes W_{3,CA^{'}})\rho_c^{\otimes2})<0$ when $c<0.406$.

\section{Appendix F: Entanglement concentration}
We compute the reduced state associated with the subsystem $AB^{'}$ after measurement. For clarity and intuitiveness, we adopt the representation method shown in Figure 1, where the first factor contains the measurement each party performs and the second factor contains the copies of the state. We have
\begin{equation}
	\begin{split}
		\xi&=\frac{1}{p_m}\frac{1}{d^3}\sum_{i,j,k,l,r,s=0}^{d-1}
		\text{Tr}_{BA^{'}}(\left(\begin{array}{cc}
			\mathbbm{1} & \mathbbm{1} \\
			(\Psi^\dagger{})^{-1}|i\rangle\langle j| & |i\rangle\langle j|(\Psi^T)^{-1}
		\end{array}\right)\left(\begin{array}{cc}
		\Psi^T|k\rangle\langle l| & |r\rangle\langle s|\Psi^{\dagger} \\
		|k\rangle\langle l|\Psi^{\dagger} & \Psi^T|r\rangle\langle s|
		\end{array}\right))\nonumber\\
		   &=\frac{1}{p_m}\frac{1}{d^3}\sum_{i,j,k,l,r,s=0}^{d-1}
		   \left(\begin{array}{cc}
		   	\Psi^T|k\rangle\langle l| & |r\rangle\langle s|\Psi^{\dagger} \\
		   	\langle j|k\rangle\langle l|i\rangle & \langle j|r\rangle\langle s|i\rangle
		   \end{array}\right)\\
		   &=\frac{1}{p_m}\frac{1}{d^3}\sum_{k,l=0}^{d-1}\Psi^T|k\rangle\langle l|\otimes|k\rangle\langle l|\Psi^{\dagger}	\\
		   &=\frac{1}{p_m}\frac{1}{d^2}|\varphi\rangle\langle\varphi|,
	\end{split}
\end{equation}
where $p_m=\text{Tr}(\mathbbm{1}_{AB^{'}}\otimes|m\rangle\langle m|_{BA^{'}}|\varphi\rangle\langle\varphi|^{\otimes2})$, which means that the final state on $AB^{'}$ is $|\varphi\rangle$ with probability $\frac{1}{d^2p_m}$.

In the above scenario, if we replace the measurement acting on $BA^{'}$ with $|M\rangle\langle M|$, where
$$
|M\rangle=\mathbbm{1}\otimes(\Psi^*\Psi^*)^{-1}|\psi^+\rangle
=(\Psi^{\dagger})^{-1}\otimes(\Psi^*)^{-1}|\psi^+\rangle,
$$
then by calculating the reduced state associated with subsystem $AB^{'}$ after measurement, we obtain
\begin{equation}
	\begin{split}
		\xi^{'}&=\frac{1}{p_m^{'}}\frac{1}{d^3}\sum_{i,j,k,l,r,s=0}^{d-1}
		\text{Tr}_{BA^{'}}(\left(\begin{array}{cc}
			\mathbbm{1} & \mathbbm{1} \\
			(\Psi^\dagger{})^{-1}|i\rangle\langle j|\Psi^{-1} & (\Psi^*)^{-1}|i\rangle\langle j|(\Psi^T)^{-1}
		\end{array}\right)\left(\begin{array}{cc}
			|k\rangle\langle l| & |r\rangle\langle s| \\
			\Psi|k\rangle\langle l|\Psi^{\dagger} & \Psi^T|r\rangle\langle s|\Psi^*
		\end{array}\right))\nonumber\\
		&=\frac{1}{p_m^{'}}\frac{1}{d^3}\sum_{i,j,k,l,r,s=0}^{d-1}
		\left(\begin{array}{cc}
			|k\rangle\langle l| & |r\rangle\langle s| \\
			\langle j|k\rangle\langle l|i\rangle & \langle j|r\rangle\langle s|i\rangle
		\end{array}\right)\\
		&=\frac{1}{p_m^{'}}\frac{1}{d^3}\sum_{k,l=0}^{d-1}|k\rangle\langle l|\otimes|k\rangle\langle l|	\\
		&=\frac{1}{p_m^{'}}\frac{1}{d^2}|\psi^+\rangle\langle\psi^+|,
	\end{split}
\end{equation}
with $p_m^{'}=\text{Tr}(\mathbbm{1}_{AB^{'}}\otimes|M\rangle\langle M|_{BA^{'}}|\varphi\rangle\langle\varphi|^{\otimes2})$. This means that the final state in $AB^{'}$ becomes a maximally entangled one with a positive probability.
\end{widetext}
\end{document}